\documentclass[12pt,twoside]{article}
\usepackage{fleqn,espcrc1}
\usepackage{epsfig}

\usepackage{graphicx}
\usepackage[figuresright]{rotating}

\newcommand{\AmS}{{\protect\the\textfont2
  A\kern-.1667em\lower.5ex\hbox{M}\kern-.125emS}}

\title{Event-by-event fluctuations in hydrodynamical description of 
heavy-ion collisions\thanks{Work supported in part by FAPESP (contract nos. 2000/04422-7 and 98/00317-2), FAPERJ (contract no.E-26/150.942/99), PRONEX 
(contract no. 41.96.0886.00) and CNPq-Brasil.}}

\author{C.E. Aguiar\address[UFRJ]{Instituto de F\'{\i}sica/UFRJ, C.P. 68528, 
        21945-970 Rio de Janeiro - RJ, Brazil},
        Y. Hama\address[USP]{Instituto de F\'{\i}sica/USP, C.P. 66318, 
        05389-970 S\~ao Paulo - SP, Brazil}, 
        T. Kodama\addressmark[UFRJ]
        and
        T. Osada\addressmark[USP]}
       
\begin{document}

\maketitle

\begin{abstract}
Effects caused by the event-by-event fluctuation of the initial conditions 
in hydrodynamical description of high-energy heavy-ion collisions are 
investigated. 
Non-negligible effects appear for several observable quantities, even for 
a fixed impact parameter $\vec b\,$. They are sensitive to the equation of 
state, being the dispersions of the observable quantities in general 
smaller when the QGP phase appears at the beginning of hydrodynamic evolution than 
when the fluid remains hadron gas during whole 
the evolution. 
\end{abstract}

\section{INTRODUCTION}

In usual hydrodynamic description of high-energy heavy-ion collisions, 
one customarily assumes some highly symmetric and smooth initial 
conditions, 
which correspond to mean distributions of velocity, temperature, energy 
density, etc., averaged over several events. However, our systems are not 
large enough, so large fluctuations are expected. What are the effects 
of 
the event-by-event fluctuation of the initial conditions? Are they sizable? 
Do they depend on the equation of state? Which are the most sensitive 
variables? These are some questions which arise regarding such 
an initial-state fluctuation, and we try to shed some light on these 
matters in the present 
study\cite{prelim}.    

\section{METHOD OF STUDY}

In order to study the problem stated above, first we generate events by 
using the NeXus event generator\cite{nexus}, from which initial 
conditions are computed at the time $\tau=1~$fm. Then, the hydrodynamic 
equations are solved, starting from these initial conditions, assuming 
some equation of state (EoS). To see the EoS dependence of the effects 
we are treating, we consider two different EoS's\cite{hung}: 
\begin{enumerate} 
\item Resonance Gas (RG): $c_s\,^2=0.2\,$;  
\item \[ 
      \mbox{QGP+RG:}\ c_s\,^2=\left\{
      \begin{array}{ll}
       0.2\,,               & \varepsilon<0.28\,\mbox{GeV/fm}^3, \\ 
       0.056/\varepsilon,   & \mbox{mixed phase}\,, \\
       1/3-4B/3\varepsilon, & \varepsilon>1.45\,\mbox{GeV/fm}^3.
      \end{array} 
      \right. 
      \]
\end{enumerate} 

The resolution of the hydrodynamic equations deserves some special care, 
since our initial conditions do not have any symmetry nor they are 
smooth. We adopt the so-called smoothed-particle hydrodynamic (SPH) 
approach\cite{sph}, first used in astrophysics and which we have 
previously adapted for heavy-ion collisions\cite{spherio}, a method 
flexible enough, giving a desired precision. The main characteristic of 
SPH is the parametrization of the flow in terms of discrete Lagrangian 
coordinates attached to small volumes (called ``particles'') with some conserved quantity. In the present work, besides the energy and momentum, 
we took the entropy as our conserved quantity. Then, its density (in the 
space-fixed frame) is parametrized as 
\begin{equation} 
  s^*({\bf x},t)=\sum_i^N \nu_i~W({\bf x}-{\bf x}_{\,i}(t);h)~, 
\end{equation} 
where 
\[ 
 \left\{ 
  \begin{array}{l} 
   W({\bf x}-{\bf x}_{\,i}(t);h)\mbox{ is a normalized kernel}; \\ 
   {\bf x}_{\,i}(t)\mbox{ is the }i\mbox{-th particle position, so the 
     velocity is }{\bf v}_{\,i}=d{\bf x}_{\,i}/dt\ ; \\  
   h~\mbox{is the smoothing scale parameter;} 
  \end{array} 
 \right. 
\] 
and we have 
\begin{equation} 
 S=\int\! d^3{\bf x}~s^*({\bf x},t)= \sum_i^N \nu_i ~. 
\end{equation} 

The equations of motion are then written as the coupled equations 
\begin{equation} 
 \frac{d}{dt}\left(\nu_i\frac{P_i+\varepsilon_i}{s_i}
                   \gamma_i{\bf v}_i\right) 
 + \sum_{j}\nu_j 
      \bigg[\frac{P_i}{{s^*_i}^2}+\frac{P_j}{{s^*_j}^2}\bigg]\, 
      {\bf\nabla}_i W({\bf x}_{\,i}-{\bf x}_{\,j};h)=0\,.\ \ \  
\end{equation} 

Following this procedure, we computed some observable quantities, 
event-by-event, for $\sqrt{s}=$130$A$GeV $Au+Au$ collisions. The results 
are presented in the next Section. 

\section{RESULTS}

\subsection{Elliptic flow coefficient $v_2$} 

Having solved the coupled equations (3), we have computed the particle 
spectra at $T=m_\pi\,$ and from which the elliptic flow coefficient 
$v_2\,$ on an event-by-event basis. 
In Figure 1, we show its distributions for a fixed impact parameter $b$, 
for the two EoS considered. As expected, $v_2$ exhibits a large 
fluctuation, which depends on the EoS. 
One should take care in looking at this Figure that our $b$ is the true 
impact parameter (not determined in the way experimentalists do), so for 
instance in the RG case, there are some events with negative $v_2\,$, 
which experimentally would not appear. 
As for the average values $<\!v_2\!>\,$, it is almost independent of 
the EoS. This is shown in Figure 2, where $< v_2 >\pm\,\delta v_2$ is 
plotted as function of the centrality and compared with data\cite{STAR}. 
It is seen that $<\!v_2\!>$ reproduces well the experimental trend, 
whereas the dispersions $\delta v_2$ are much wider than the experimental 
errors. As for the EoS dependence, $\delta v_2$ is smaller when QGP is 
produced. 

\begin{figure}[htb]
\begin{minipage}[t]{80mm}
{\epsfysize=6.5cm \epsfig{file=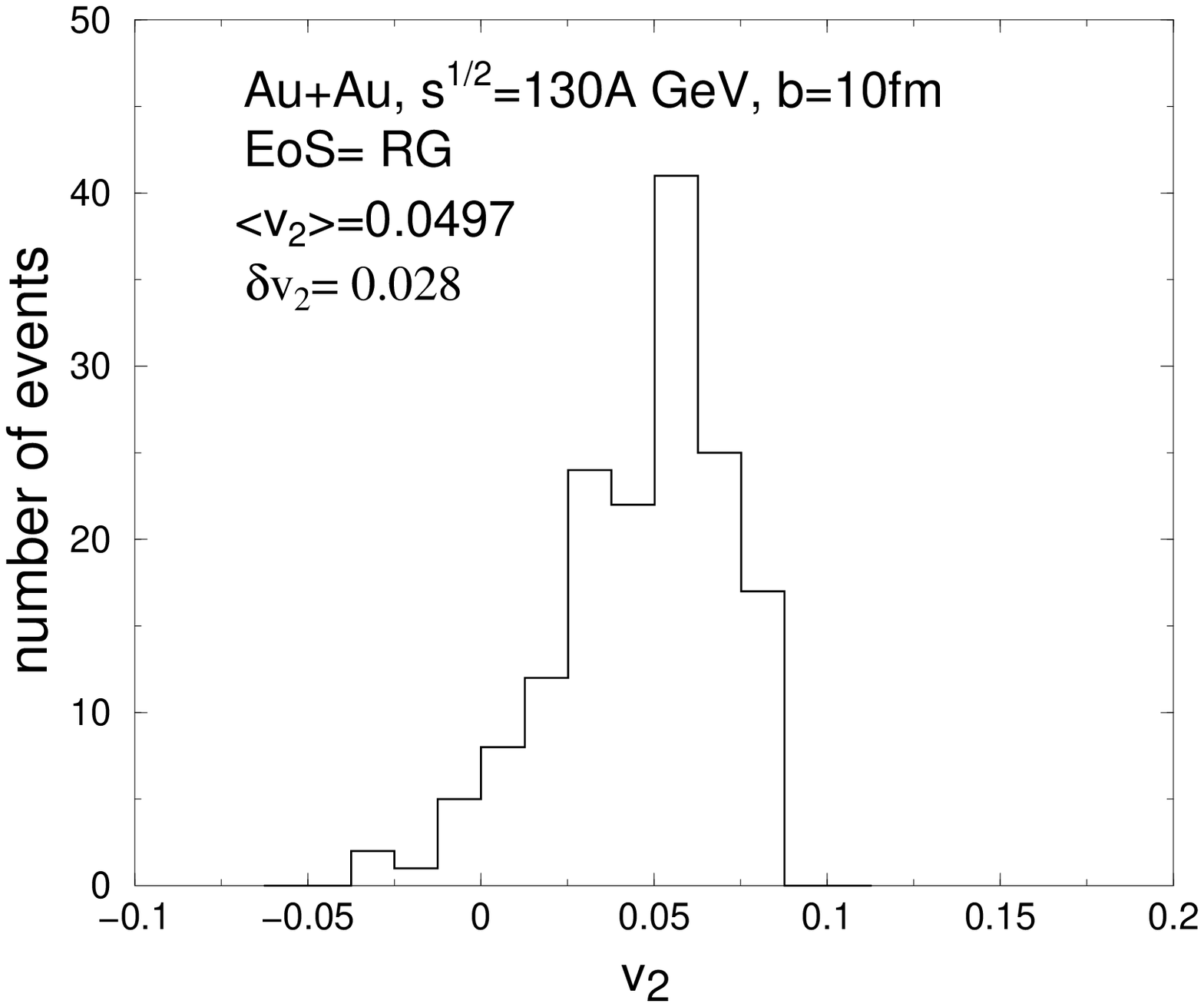}}
\end{minipage}
\hspace{\fill}
\begin{minipage}[t]{75mm}
{\epsfysize=6.5cm \epsfig{file=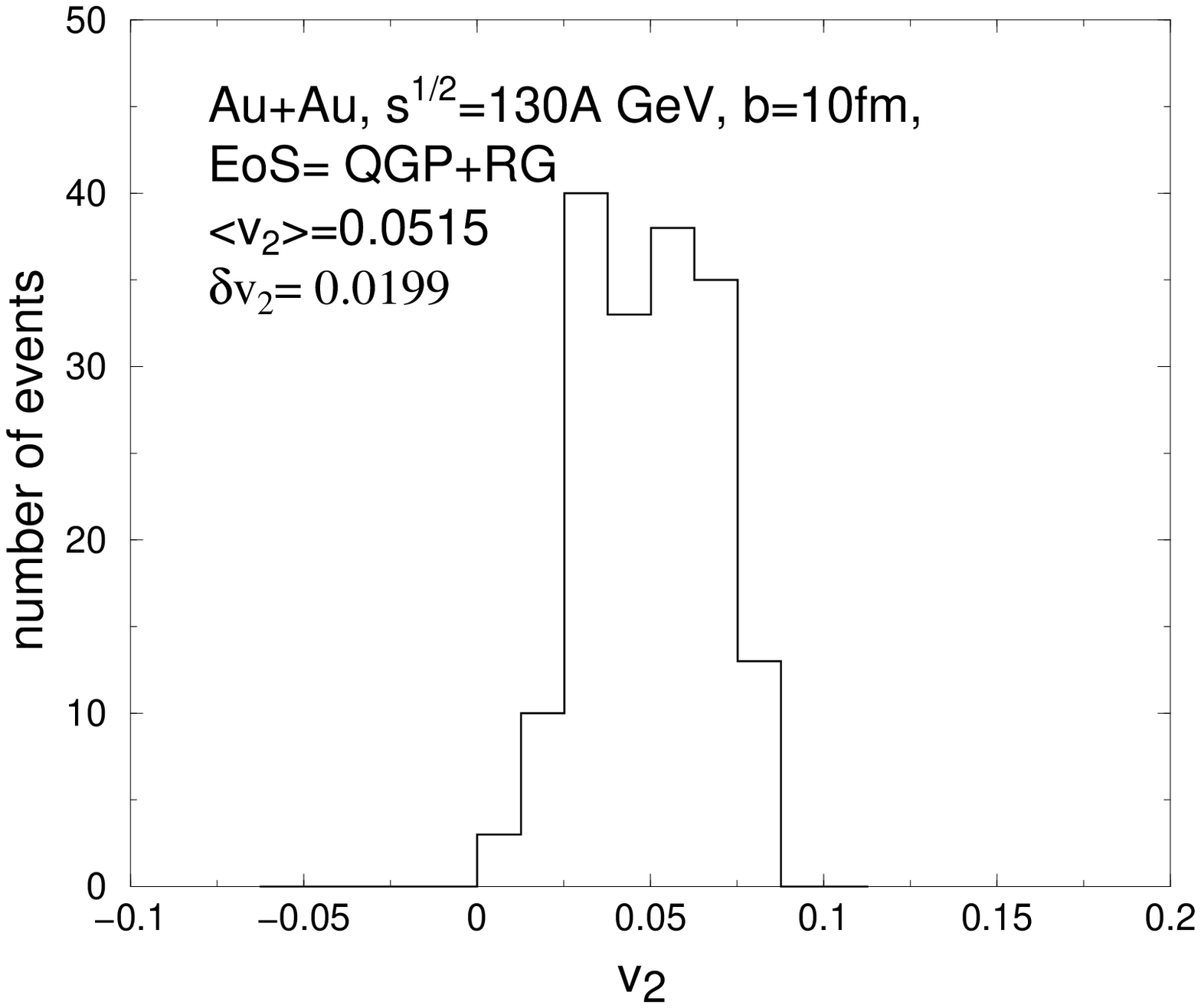}}
\end{minipage}
\vspace*{-1.cm}
\caption{Distribution of elliptic-flow coefficients $v_2$ at $b=10~$fm for 
two EoS.} 
\end{figure} 

\medskip

\begin{figure}[htb] 
\begin{minipage}[t]{80mm}
{\epsfysize=6.2cm \epsfig{file=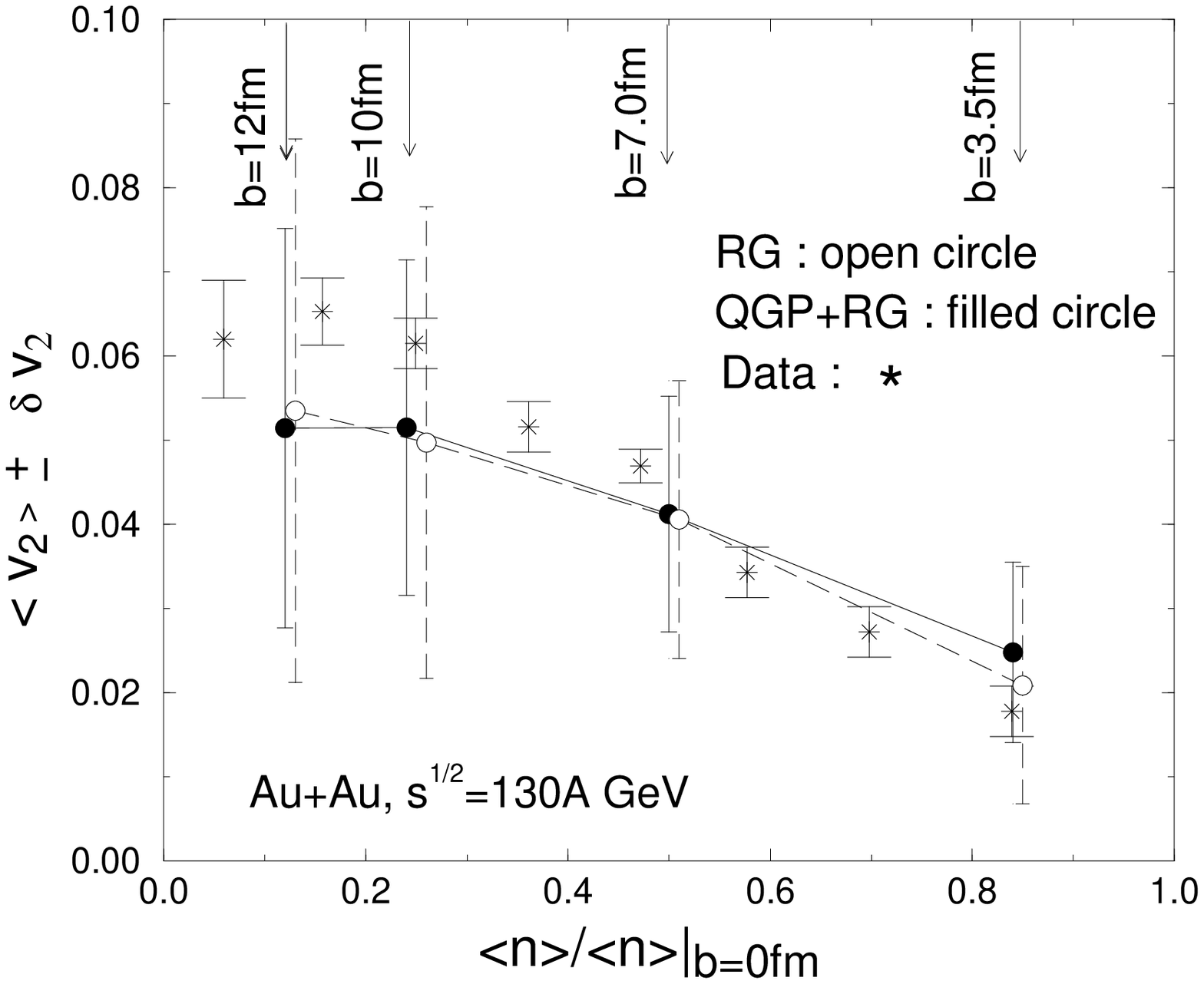}} 
\vspace*{-1.cm} 
\caption{EoS dependence of $< v_2 >\pm\,\delta v_2$ as function of the 
centrality, compared with data\cite{STAR}. }
\end{minipage}
\hspace{\fill}
\begin{minipage}[t]{75mm}
{\epsfysize=6.2cm \epsfig{file=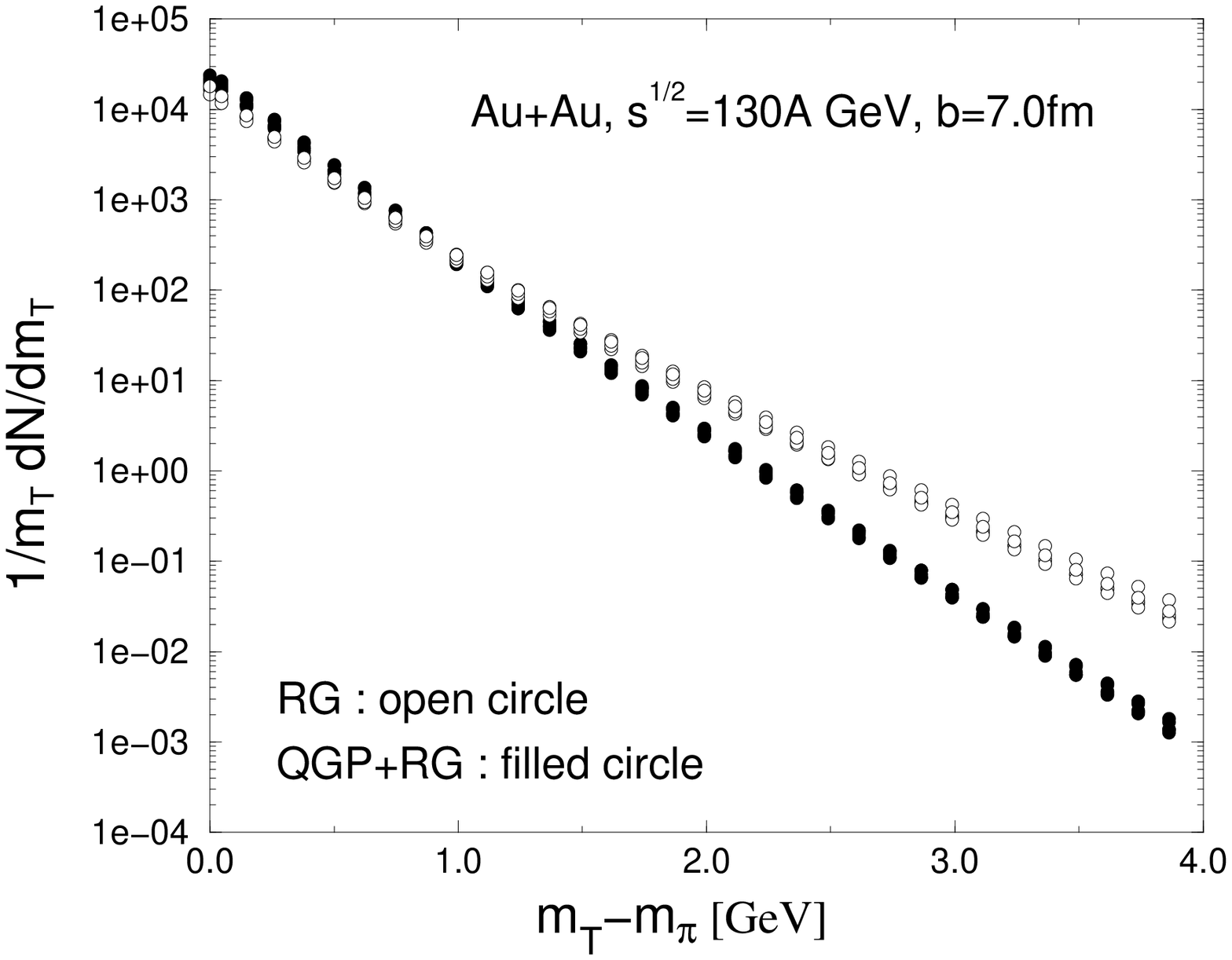}} 
\vspace*{-1.cm} 
\caption{Pion $m_T$ spectra for 5 events ($b=7.0~$fm). }
\end{minipage}
\end{figure}

\subsection{{\boldmath $m_T$} distributions} 

In Figure 3, we show the $m_T$ distributions for 5 events. As expected, 
$m_T$ distributions are in general steeper when QGP is produced. As for 
fluctuations, the resultant fluctuation in $m_T$ spectrum (or in the 
slope parameter $\delta\tilde{T}$) is very small. 

\subsection{Multiplicity fluctuation in the mid-rapidity region} 

Table 1 summarizes the results of our study on multiplicity fluctuation 
in the mid-rapidity region. It is seen that {\it i)} as $b\rightarrow0$, 
$<n_\pi>$ becomes much larger with the QGP EoS; {\it ii)} $\delta n_\pi$ 
shows the same tendency in this limit; {\it iii)} As for the ratio 
$\delta n_\pi/\!<n_\pi>$, it is not sensitive to the EoS. 

\begin{table}[htb] 
\caption{EoS dependence of the multiplicity fluctuation in two different 
 rapidity intervals $-\Delta y<y<+\Delta y$, as function of the impact 
 parameter $b$.} 
\label{table:1} 
\newcommand{\m}{\hphantom{$-$}}
\newcommand{\cc}[1]{\multicolumn{1}{c}{#1}}
\renewcommand{\arraystretch}{1.2} 
\newcommand{\lw}[1]{\smash{\lower2.0ex\hbox{#1}}} 
\renewcommand{\arraystretch}{1.2} 
 
\begin{tabular}{rcr|rcc|rrc}
\hline
\lw{$b~$[fm]$\!\!$} & $\ $ \lw{EoS}$\ $ & \# of$\ \ $ 
&\multicolumn{3}{c|}{$\Delta y$=1.875} &
\multicolumn{3}{c}{$\Delta y$=3.00} \\ 
\cline{4-9} & & $\ \ $events$\ $  
&$\!<\!\!n\!\!>\ $&$\delta n$~&$\delta n/\!\!<\!\!n\!\!>$ 
&$\!<\!\!n\!\!>\ $&$\delta n\ \ $~&$\delta n/\!\!<\!\!n\!\!>$ \\ 
\hline
\lw{3.5} & RG  & 44$\ \ \ $ & $\ $1029.7& $\ $46.2$\ $ & 0.045 
         & $\ \ $1623.2     & 68.7      & $\ $0.042$\ $ \\
         & QGP & 38$\ \ \ $ & 1553.0& 80.9 & 0.052 &2544.6& 129.0 &0.051\\
\hline 
\lw{7.0} & RG  & 55$\ \ \ $ & 613.3 & 49.5 & 0.081 &977.4 & 71.6  &0.073\\
         & QGP & 58$\ \ \ $ & 926.1 & 81.1 & 0.087 &1530.5& 123.7 &0.081\\
\hline 
\lw{10.0} & RG &166$\ \ \ $ & 312.8 & 43.0 & 0.137 &506.1 &  65.8 &0.130\\
         & QGP &180$\ \ \ $ & 437.5 & 66.5 & 0.151 &740.7 & 103.9 &0.140\\
\hline 
\lw{12.0} & RG & 79$\ \ \ $ & 162.8 & 35.6 & 0.219 & 268.9 & 56.2 &0.209\\
         & QGP &100$\ \ \ $ & 220.1 & 52.8 & 0.240 & 379.8 & 85.2 &0.224\\
\hline 
  \end{tabular}

\end{table}

\section{CONCLUSIONS AND OUTLOOK} 

The present study shows that the effects of the event-by-event fluctuation 
of the initial conditions in hydrodynamics are sizable and should be 
considered in data analyses. They do depend on the equation of state. 
Among the quantities examined here, $\delta v_2$ is the most sensitive to the equation of state. 

In the present work, many important factors have not been considered:  
baryon-number conservation, strangeness production, resonance decays, 
continuous emission effects, spectators, etc., which should indeed taken 
into account in order to get more precise results. 
Especially, use of the same procedure for the determination of the 
centrality as used by experimentalists, as in\cite{STAR}, will make the 
results more directly comparable with data.  
In any event, we believe that the effects we studied will be present 
and will be sizable, even with these improvements.

\end{document}